\documentclass{article}

\usepackage{microtype}
\usepackage{graphicx}
\usepackage{subfigure}
\usepackage{booktabs} 
\usepackage{amsmath,amsfonts,amssymb}
\usepackage{natbib}
\usepackage{hyperref}
\usepackage{tikz}
\def\checkmark{\tikz\fill[scale=0.4](0,.35) -- (.25,0) -- (1,.7) -- (.25,.15) -- cycle;} 
\usepackage{fullpage}

\usepackage{algorithm}
\usepackage{algorithmic}

\newcommand{\x}{\mathbf{x}}
\newcommand{\w}{\mathbf{w}}

\usepackage[textsize=footnotesize,color=green!40]{todonotes}

\usepackage{authblk}
\title{TAPAS: Tricks to Accelerate (encrypted) Prediction As a Service}

\newif\ifdraft

\usepackage{setspace}

\newcommand{\vone}{\mathbf{1}}
\newcommand{\vb}{\mathbf{b}}

\newcommand{\sign}{\texttt{sign}}
\newcommand{\integers}{\mathbb{Z}}

\drafttrue

\date{}
\begin{document}

\author[1,2]{Amartya Sanyal\thanks{amartya.sanyal@cs.ox.ac.uk}}
\author[2,3]{Matt J. Kusner\thanks{mkusner@turing.ac.uk}}
\author[2,3]{Adri{\`a} Gasc{\'o}n\thanks{agascon@turing.ac.uk}}
\author[1,2]{Varun Kanade\thanks{varunk@cs.ox.ac.uk}}

\affil[1]{University of Oxford, Oxford, UK}
\affil[2]{The Alan Turing Institute, London, UK}
\affil[3]{University of Warwick, Coventry, UK}

\maketitle
\begin{abstract}
	Machine learning methods are widely used for a variety of prediction problems. \emph{Prediction as a service} is a paradigm in which service providers with technological expertise and computational resources may perform predictions for clients. However, data privacy severely restricts the applicability of such services, unless measures to keep client data private (even from the service provider) are designed. Equally important is to minimize the amount of computation and communication required between client and server. Fully homomorphic encryption offers a possible way out, whereby clients may encrypt their data, and on which the server may perform arithmetic computations. The main drawback of using fully homomorphic encryption is the amount of time required to evaluate large machine learning models on encrypted data. We combine ideas from the machine learning literature, particularly work on binarization and sparsification of neural networks, together with algorithmic tools to speed-up and parallelize computation using encrypted data.
\end{abstract}

\section{Introduction}
\label{sec:intro}
%
Applications using machine learning techniques have exploded during the recent
years, with ``deep learning'' techniques being applied on a wide variety of
tasks that had hitherto proved challenging. Training highly accurate machine
learning models requires large quantities of (high quality) data, technical
expertise and computational resources. An important recent paradigm is
\emph{prediction as a service}, whereby a service provider with expertise and resources
can make predictions for clients. However, this approach requires trust between
service provider and client; there are several instances where clients may be
unwilling or unable to provide data to service providers due to privacy concerns. 
Examples include assisting in medical diagnoses \citep{K:2001,BSHLKPRK:2017},
detecting fraud from personal finance data \citep{GR:1994}, and detecting
online communities from user data \citep{F:2010}. The ability of a service provider
to predict on encrypted data can alleviate concerns of data leakage. 

The framework of fully homomorphic encryption (FHE) is ideal for this paradigm.
Fully homomorphic encryption schemes support arbitrary computations to be performed
directly on encrypted data without prior decryption. The first fully homomorphic encryption system 
was developed just 10 years ago by \citet{Gen:2009}, after
being an open question for 30 years~\citep{RAD:1978}. Since then several other schemes
have been proposed \cite{GHS:2012,GSW:2013,BV:2014a,DM:2015,CGGI:2016}.  However, 
without significant changes to machine learning models and improved algorithmic 
tools, homomorphic encryption does not scale to real-world machine learning applications.

Indeed, already there have been several recent works trying to accelerate predictions of machine 
learning models on fully homomorphic encrypted data. In general, the approach has been to
approximate all or parts of a machine learning model
to accommodate the restrictions of an FHE framework.
Often, certain kind of FHE schemes 
are preferred 
because they allow for ``batched'' parallel encrypted computations, called
SIMD operations \cite{SV:2014}. This technique is exemplified by the CryptoNets model 
\citep{G-BDL+:2016}. While these models allow for high-throughput (via SIMD), they are not particularly
suited for the prediction as a service framework for individual users, as single predictions are
slow. Further, because they employ a leveled homomorphic encryption scheme, they are unable to perform many nested
multiplications, a requirement for state-of-the-art deep learning models \cite{HZRS:2016,HZWV:2017}.

Our solution demonstrates that existing work on Binary Neural Networks (BNNs) \citep{KS:2015,CHSEB:2016} can be adapted to produce efficient and highly accurate
predictions on encrypted data.
We show that a recent 
FHE encryption scheme~\cite{CGGI:2016}
which only supports operations on binary data can be leveraged to
compute all of the operations of BNNs. To do so, we develop specialized circuits
for fully-connected, convolutional, and batch normalization layers \cite{IS:2015}.
Additionally we design tricks to sparsify encrypted computation that reduce computation
time even further.

Most similar to our work is \citet{BMM+:2017} who use neural networks with 
signed integer weights and binary activations to perform encrypted prediction. However,
this model is only evaluated on MNIST, with modest accuracy results,
and the encryption scheme parameters depend on the structure of the model, potentially
requiring clients to re-encrypt their data if the service provider updates their model.
Our framework allows the service provider to update
their model at anytime, and allows one to use binary neural networks of~\citet{CHSEB:2016} 
which, in particular, achieve high accuracy on MNIST ($99.04\%$).
Another closely related work is \citet{MLH:2018} who design encrypted adder and multiplier 
circuits so that they can implement machine learning models on integers. This can be seen as
complementary to our work on binary networks: while they achieve improved accuracy because
of greater precision, they are less efficient than our methods (however on MNIST we achieve the same accuracy with a $29\times$ speedup, via our sparsification and parallelization tricks).

\paragraph{Private training.}
In this work, we do not address the question of training machine learning models with encrypted data. There has been some recent work in this area \citep{HHI-LNPST:2017,AHWM:2017}. However, as of now it appears possible only to train very small models using fully homomorphic encryption. We leave this for future work. 

\subsection{Our contributions}
\label{ssec:contrib}
In this work, our focus is on achieving speed-ups when using complex models on
fully homomorphic encrypted data. In order to achieve these speed-ups, we
propose several methods to modify the training and design of neural networks,
as well as algorithmic tricks to parallelize and accelerate computation on
encrypted data:
\begin{itemize}
	\item We propose two types of circuits for performing inner products between
		unencrypted and encrypted data: reduce tree circuits and sorting
		networks. We give a runtime comparison of each method.
	\item We introduce an easy trick, which we call the \emph{+1 trick} to
		sparsify encrypted computations.
	\item We demonstrate that our techniques are easily parallelizable and we report 
		timing for a variety of computation settings on real world datasets,
		alongside classification accuracies.
\end{itemize}

\section{Encrypted Prediction as a Service}
\label{sec:problem-definition}

\begin{figure}[t]
    \centerline{\includegraphics[width=0.5\linewidth]{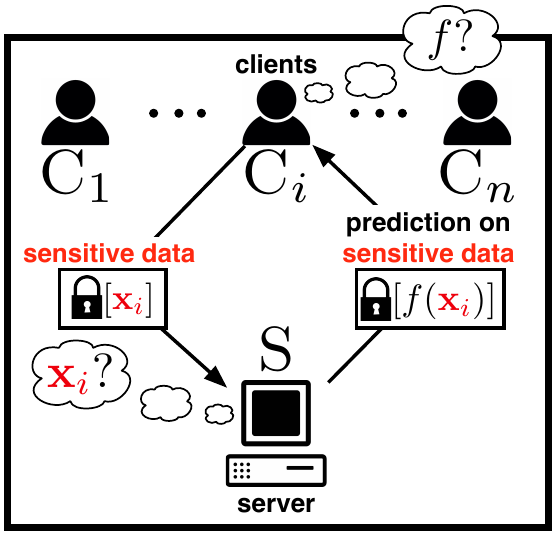}}
    \vspace{-2ex}
  \caption{Encrypted prediction as a service. }
  \label{figure.service}
\end{figure}

In this section we describe our \emph{Encrypted Prediction as a Service (EPAAS)} paradigm. We then detail our privacy and computational guarantees. Finally, we discuss how different related work is suited to this paradigm and propose a solution.

In the EPAAS setting we have any number of clients, say $C_1, \ldots, C_n$ that have data $\x_1,\ldots,\x_n$. The clients would like to use a highly-accurate model $f$ provided by a server $S$ to predict some outcome. In cases where data $\x$ is not sensitive there are already many solutions for this such as BigML, Wise.io, Google Cloud AI, Amazon Machine Learning, among others. However, if the data is sensitive so that the clients would be uncomfortable giving the raw data to the server, none of these systems can offer the client a prediction. 


\subsection{Privacy and computational guarantees}
If data $\x$ is sensitive (e.g., $\x$ may be the health record of client $C$, and $f(\x)$ may be the likelihood of heart disease), then we would like to have the following privacy guarantees:
\begin{enumerate}
\item[P1.] Neither the server $S$, or any other party, learn anything about client data $\x$, other than its size {\em (privacy of the data)}.
\item[P2.] Neither the client $C$, or any other party, learn anything about model $f$, other than the prediction
  $f(x)$ given client data $\x$ (and whatever can be deduced from it) {\em (privacy of the model)}. 
\end{enumerate}
Further, the main attraction of EPAAS is that the client is involved as little as possible. More concretely, we wish to have the following computational guarantees:
\begin{enumerate}
\item[C1.] No external party is involved in the computation.
\item[C2.] The rounds of communication between client and server should be limited to $2$ (send data \& receive prediction).
\item[C3.] Communication and computation at the client side
  should be independent of model $f$. In particular,
  (i) the server should be able to update $f$ without communicating with any client,
  and (ii) clients should not need to be online during the computation of $f(\x)$.
\end{enumerate}
Note that these requirements rule out protocols with preprocessing stages or
that involve third parties.
Generally speaking, a satisfactory solution based on FHE would proceed as follows:
(1) a client generates encryption parameters, encrypts their data $\x$ using
the private key, and sends the resulting encryption $\tilde{\x}$, as well as the
public key to the server. (2) The server evaluates $f$ on $\tilde{\x}$ leveraging the homomorphic
properties of the encryption, to obtain an encryption $\tilde{f}(\x)$
without learning anything whatsoever about $\x$, and sends $\tilde{f}(\x)$
to the client. (3) Finally, the client decrypts and recovers the prediction
$f(\x)$ in the clear. A high level depiction of these steps is shown in Figure~\ref{figure.service}. 

\begin{table*}[!htb]
\centering
	\caption{Privacy and computational guarantees of existing methods for sensitive data classification. \label{table.existing}}
\begin{tabular}{c|cc|cccc} 
\hline
 &  \multicolumn{2}{c}{\bf Privacy} & \multicolumn{4}{|c}{\bf Computation} \\
Prior Work  &  {P1} & {P2} & {C1} & {C2} & {C3(i)} & {C3(ii)} \\
\hline
CryptoNets~\cite{G-BDL+:2016} & \checkmark & - & \checkmark & \checkmark & - & \checkmark \\
\hline
\cite{CWMMP:2017} & \checkmark & - & \checkmark & \checkmark & - & \checkmark \\
\hline
\cite{BMM+:2017} & \checkmark & \checkmark & \checkmark & \checkmark & - & \checkmark \\
\hline
\begin{tabular}{@{}c@{}}MPC~ \cite{MZ:2017,LJL+:2017,RRK:2017} \\ \cite{RWT+:2017,CG-BLLR:2017,JVC:2018}\end{tabular} & - & \checkmark & \checkmark & - & - & - \\
\hline
\cite{MLH:2018}, Ours & \checkmark & \checkmark & \checkmark & \checkmark & \checkmark & \checkmark \\
\hline
\end{tabular}
\end{table*}

\subsection{Existing approaches}
Table~\ref{table.existing} describes whether prior work satisfy the above privacy and computational guarantees. First, note that Cryptonets~\cite{G-BDL+:2016} violates C3(i) and P2. This is because the clients would have to generate
parameters for the encryption according to the structure of $f$, so we are able to make inferences about the model (violating P2) and the client is not allowed to change the model $f$ without telling the client (violating C3(i)). The same holds for the work of \citet{CWMMP:2017}. The approach of \citet{BMM+:2017} requires the server to calibrate the parameters of the encryption scheme according to the
magnitude of intermediate values, thus C3(i) is not necessarily satisfied. Closely related to our work is that of \citet{MLH:2018} which satisfies our privacy and computational requirements. We will show that our method is significantly faster than this method, with very little sacrifice in accuracy.

\paragraph{Multi-Party Computation (MPC).}
It is important to distinguish between
approaches purely based on homomorphic encryption (described above), and those
involving Multi-Party Computation (MPC) techniques, such as~\cite{MZ:2017,LJL+:2017,RRK:2017,RWT+:2017,CG-BLLR:2017,JVC:2018}. 
While generally MPC approaches are faster, they crucially
rely on all parties being involved in the whole computation,
which is in conflict with requirement C3(ii).
Additionally, in MPC the structure of the computation is public to both parties,
which means that the server would have to communicate basic information such as
the number of layers of $f$. This is conflict with requirements P1,
C2, and C3(i).

In this work, we propose to use a very tailored homomorphic encryption technique to guarantee all privacy and computational requirements. In the next section we give background on homomorphic encryption. Further, we motivate the encryption protocol and the machine learning model class we use to satisfy all guarantees.



\section{Background}
\label{sec:prelim}

All cryptosystems define two functions: 1. an encryption function $\mathcal{E}(\cdot)$ that maps data (often called \emph{plaintexts}) to encrypted data (\emph{ciphertexts}); 2. a decryption function $\mathcal{D}(\cdot)$ that maps ciphertexts back to plaintexts. In public-key cryptosystems, to evaluate the encryption function $\mathcal{E}$, one needs to hold a public key $k_\textsc{pub}$, so the encryption of data $x$ is $\mathcal{E}(x, k_\textsc{pub})$. Similarly, to compute the decryption function $\mathcal{D}(\cdot)$ one needs to hold a secret key $k_\textsc{sec}$ which allows us to recover: $\mathcal{D}(\mathcal{E}(x, k_\textsc{pub}),k_\textsc{sec}) = x$.

A cryptosystem is \emph{homomorphic} in some operation $\blacksquare$ if it is possible to perform another (possibly different) operation $\square$ such that: $\mathcal{E}(x, k_\textsc{pub}) \; \square \; \mathcal{E}(x, k_\textsc{pub}) = \mathcal{E}(x \; \blacksquare \; y, k_\textsc{pub})$. Finally, in this work we assume all data to be binary $\in \{0,1\}$. For more detailed background on FHE beyond what is described below, see the excellent tutorial of \citet{H:2017}.

\subsection{Fully Homomorphic Encryption}
\label{ssec:fhe}
In 1978, cryptographers posed the question: \emph{Does an encryption scheme exist that allows one to perform arbitrary computations on encrypted data?} The implications of this, called a \emph{Fully homomorphic encryption} (FHE) scheme, would enable clients to send computations to the cloud
while retaining control over the secrecy of their data.
This was still an open problem however 30 years later. Then, in 2009, a cryptosystem \citep{Gen:2009} was devised that could, in principle, perform such computations on encrypted data. Similar to previous approaches, in each computation, noise is introduced into the encrypted data. And after a certain number of computations, the noise grows too large so that the encryptions can no longer be decrypted. 
The key innovation was a technique called \emph{bootstrapping}, which allows one to reduce the noise to its original level without decrypting. 
That result constituted a massive breakthrough, as it established, for the first time, a fully homomorphic encryption scheme \citep{Gen:2009}. 
Unfortunately, the original bootstrapping procedure was highly impractical. 
\begin{figure*}[t!]
    \centerline{\includegraphics[width=0.75\textwidth]{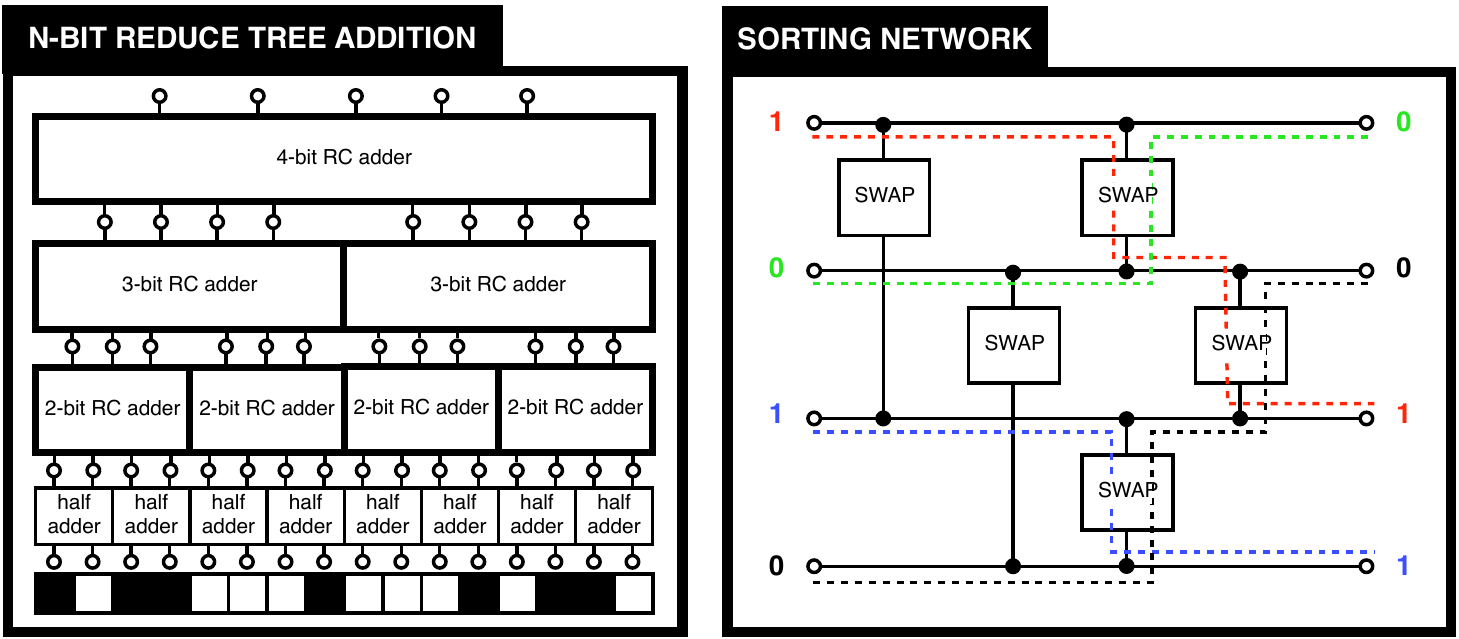}}
    \vspace{-2ex}
  \caption{Binary circuits used for inner product: reduce tree (\emph{Left}) and sorting network (\emph{Right}). RC is short for ripple-carry.}
  \label{figure.binary_ops}
\end{figure*}
Consequently, much of the research since the first FHE scheme has been devoted to reducing the growth of noise so that the scheme never has to perform bootstrapping. 
Indeed, even in recent FHE schemes bootstrapping is slow 
(roughly six minutes in a highly-optimized implementation of a recent popular scheme \citep{HS:2015}) and bootstrapping many times increases the memory requirements of encrypted data. 

\subsubsection{Encrypted Prediction with leveled HE}
Thus, one common technique to implement encrypted prediction was to take an existing ML algorithm and approximate it with as few operations as possible, in order to never have to bootstrap. This involved careful parameter tuning to ensure that the security of the encryption scheme was sufficient, that it didn't require too much memory, and that it ran in a reasonable amount of time. One prominent example of this is Cryptonets \citep{G-BDL+:2016}.

\subsubsection{Encrypted Prediction with FHE}
Recent developments in cryptography call for rethinking this approach. \citet{DM:2015} devised a scheme that that could bootstrap a single Boolean gate in under one second with reduced memory. Recently, \citet{CGGI:2016} introduced optimizations implemented in the TFHE library, which further reduced bootstrapping of to under 0.1 seconds. In this paper, we demonstrate that this change has a huge impact on designing encrypted machine learning algorithms. Specifically, encrypted computation is now modular: the cost of adding a few layers to an encrypted neural network is simply the added cost of each layer in isolation. This is particularly important as recent developments in deep learning such as Residual Networks \citep{HZRS:2016} and Dense Networks \citep{HZWV:2017} have shown that networks with many layers are crucial to achieve state-of-the-art accuracy.

\subsection{Binary Neural Networks}
\label{ssec:binary}
The cryptosystem that we will use in this paper, TFHE, is however restricted to computing binary operations. We note that, concurrent to the work that led to TFHE, was the development of neural network models that perform binary operations between binary weights and binary activations. These models, called Binary Neural Networks (BNNs), were first devised by \citet{KS:2015,CHSEB:2016}, and were motivated by the prospect of training and testing deep models on limited memory and limited compute devices, such as mobile phones.

\paragraph{Technical details.}
We now describe the technical details of binary networks that we will aim to replicate on encrypted data. In a \emph{Binary Neural Network} (BNN) every layer maps a binary input $\mathbf{x} \in \{-1,1\}^{d}$ to a binary output $\mathbf{z} \in \{-1,1\}^p$ using a set of binary weights $\mathbf{W} \in \{-1,1\}^{(p,d)}$ and a binary activation function $\texttt{sign}(\cdot)$ that is $1$ if $x \geq 0$ and $-1$ otherwise. 
 Although binary nets don't typically use a bias term, applying batch-normalization~\citep{IS:2015} when evaluating the model it means that a bias term $\vb \in \integers^p$ may need to be added before applying the activation function (cf. Sec.~\ref{subsubsec:bn}). Thus, when evaluating the model, a fully connected layer in a BNN implements the following transformation $\mathbf{z} := \texttt{sign}(\mathbf{W}\mathbf{x} + \vb)$. From now on we will call all data represented as $\{-1,1\}$ \emph{non-standard binary} and data represented as $\{0,1\}$ as \emph{binary}. \citet{KS:2015,CHSEB:2016} were the first to note that the above inner product nonlinearity in BNNs could be implemented using the following steps:

\begin{enumerate}
	\item Transform data and weights from non-standard binary to binary:
		$\mathbf{w},\mathbf{x} \rightarrow
		\overline{\mathbf{w}},\overline{\mathbf{x}}$ by replacing $-1$ with $0$.
		n
	\item Element-wise multiply by applying the logical operator
		\texttt{XNOR}$(\overline{\mathbf{w}},\overline{\mathbf{x}})$ for each
		element of $\overline{\mathbf{w}}$ and $\overline{\mathbf{x}}$.  
	\item Sum result of previous step by using \texttt{popcount} operation
		(which counts the number of 1s), call this $S$.
	\item If the bias term is $b$, check if $2 S \geq d - b$, if so the
		activation is positive and return $1$, otherwise return $-1$.
\end{enumerate}

Thus we have that,
\begin{align*}
	z_i = \sign(2 \cdot \texttt{popcount}(\texttt{XNOR}(\overline{\mathbf{w}}_{i},\overline{\mathbf{x}})) - d + b)
\end{align*}

\paragraph{Related binary models.}
Since the initial work on BNNs there has been a wealth of work on binarizing, ternarizing, and quantizing neural networks \citet{CWTWC:2015,CBD:2015,HMD:2016,HCSEB:2016,ZHMD:2016,CWMMP:2017,CHZX:2017}. Our approach is currently tailored to methods that have binary activations and we leave the implementation of these methods on encrypted data for future work.



\section{Methods}
\label{sec:methods}
In this work, we make the observation that BNNs can be run on encrypted data by designing circuits in TFHE for computing their operations. 
In this section we consider Boolean circuits that operate on encrypted data and
unencrypted weights and biases. We show how these circuits allow us to
efficiently implement the three main layers of binary neural networks: fully
connected, convolutional, and batch-normalization. We then show how a simple
trick allows us to sparsify our computations.  Our techniques can be easily
parallelized.  During the evaluation of a circuit, gates at the same level in
the tree representation of the circuit can be evaluated in parallel. Hence,
when implementing a function, ``shallow'' circuits are preferred in terms of
parallelization. While parallel computation was often used to justify employing
second generation FHE techniques---where parallelization comes from ciphertext
packing---we show in the following section that our techniques create dramatic
speedups for a state-of-the-art FHE technique. We emphasize that a key
challenge is that we need to use \emph{data oblivious} algorithms (circuits)
when dealing with encrypted data as the algorithm never discovers the actual
value of any query made on the data.


\subsection{Binary OPs}
\label{ssec:binaryops}
The three primary circuits we need are for the following tasks: 1. computing the inner product; 2. computing the binary activation function (described in the previous section) and; 3. dealing with the bias.
\begin{algorithm}[H]
  \caption{Comparator}
  {\bf Inputs:}~~Encrypted $\mathbb{B}[\tilde{S}]$, unencrypted $\mathbb{B}[(d-b)/2]$, size $d$ of $\mathbb{B}[(d-b)/2]$,$\mathbb{B}[\tilde{S}]$ \\
  {\bf Output:}~~Result of $2\tilde{S} \geq d - b$
  \begin{algorithmic}[1]
  \label{alg:compare}
  	\STATE $o = 0$
  	\FOR{$i = 1,\ldots,d$}
    	\IF{$\mathbb{B}[(d-b)/2]_i = 0$}
    		\STATE $o = \textsc{MUX}(\mathbb{B}[\tilde{S}]_i, \tilde{1}, o)$
    	\ELSE
    		\STATE $o = \textsc{MUX}(\mathbb{B}[\tilde{S}]_i, o, \tilde{0})$
    	\ENDIF
    \ENDFOR
    \STATE {\bf Return:} $o$
  \end{algorithmic}
\end{algorithm}
\subsubsection{Encrypted inner product}
As described in the previous section, BNNs can speed up an inner product by computing \textsc{XNOR}s (for element-wise multiplication) followed by a \textsc{popcount} (for summing). In our case, we compute an inner product of size $d$ by computing \textsc{XNOR}s element-wise between $d$ bits of encrypted data and $d$ bits of unencrypted data, which results in an encrypted $d$ bit output. To sum this output, the \textsc{popcount} operation is useful when weights and data are unencrypted because \textsc{popcount} is implemented in the instruction set of Intel and AMD processors, but when dealing with encrypted data we simply resort to using shallow circuits. We consider two circuits for summation, both with sublinear depth:
a reduce tree adder and a sorting network.


\paragraph{Reduce tree adder.}
We implement the sum using a binary tree of half and ripple-carry (RC) adders organized into a reduction tree, as shown in Figure~\ref{figure.binary_ops} (\emph{Left}). All these structures can be implemented to run on encrypted data because TFHE allows us to compute \textsc{XNOR}, \textsc{AND}, and \textsc{OR} on encrypted data. The final number returned by the reduction tree $\tilde{S}$ is the binary representation of the number of $1$s resulting from the $\textsc{XNOR}$, just like \textsc{popcount}. Thus, to compute the BNN activation function \sign$(\cdot)$ we need to check whether $2\tilde{S} \geq d - b$, where $d$ is the number of bits in $\tilde{S}$ and $b$ is the bias. Note that if the bias is zero we simply need to check if $\tilde{S} \geq d/2$. To do so we can simply return the second-to-last bit of $\tilde{S}$. If it is $1$ then $\tilde{S}$ is at least $d/2$. If the bias $b$ is non-zero (because of batch-normalization, described in Section~\ref{subsubsec:bn}), we can implement a circuit to perform the check $2\tilde{S} \geq d - b$. The bias $b$ (which is available in the clear) may be an integer as large as $\tilde{S}$. Let $\mathbb{B}[(d-b)/2]$, $\mathbb{B}[\tilde{S}]$ be the binary representations of $b$ and $\tilde{S}$. Algorithm~\ref{alg:compare} describes a comparator circuit that returns an encrypted value of $1$ if the above condition holds and (encrypted) $0$ otherwise (where $\textsc{MUX}(s,a,b)$ returns $a$ if $s=1$ and $b$ otherwise). As encrypted operations dominate the running time of our computation, in practice this computation essentially corresponds to evaluating $d$ MUX gates. This gate has a dedicated implementation in TFHE, which results in a very efficient comparator in our setting.


\paragraph{Sorting network.}
We do not technically care about the sum of the result of the element-wise $\textsc{XNOR}$ between $\bar{\mathbf{w}}$ and $\bar{\mathbf{x}}$. In fact, all we care about is if the result of the comparison: $2\tilde{S} \geq d - b$. Thus, another idea is to take the output of the (bitwise) $\textsc{XNOR}$ and sort it. Although this sorting needs to be performed over encrypted data, the rest of the computation does not require any homomorphic operations; after sorting we hold a sequence of encrypted $1$s, followed by encrypted $0$s. To output the correct value, 
we only need to select one the (encrypted) bit in the correct position and return it.
If $b=0$ we can simply return the encryption of the central bit in the sequence; indeed, if the central bit is $1$, then there are more $1$s than $0$s and thus $2\tilde{S} \geq d$ and we return $1$. If $b\neq0$ we need to offset the returned index by $b$ in the correct direction depending on the sign of $b$. In order to sort the initial array we implement a sorting network, shown in Figure~\ref{figure.binary_ops} (\emph{Right}). The sorting network is a sequence of swap gates between individuals bits, where $\textsc{SWAP}(a,b) = (\textsc{OR}(a,b), \textsc{AND}(a,b))$. Note that if $a \geq b$ then $\textsc{SWAP}(a,b) = (a,b)$, and otherwise is $(b,a)$. More specifically, we implement Batcher's sorting network~\cite{batcher_sorting_1968}, which consists of $O(n \log^2(n))$ swap gates, and has depth $O(\log^2(n))$.

\begin{figure}[H]
  \centering
   \includegraphics[width=0.5\linewidth]{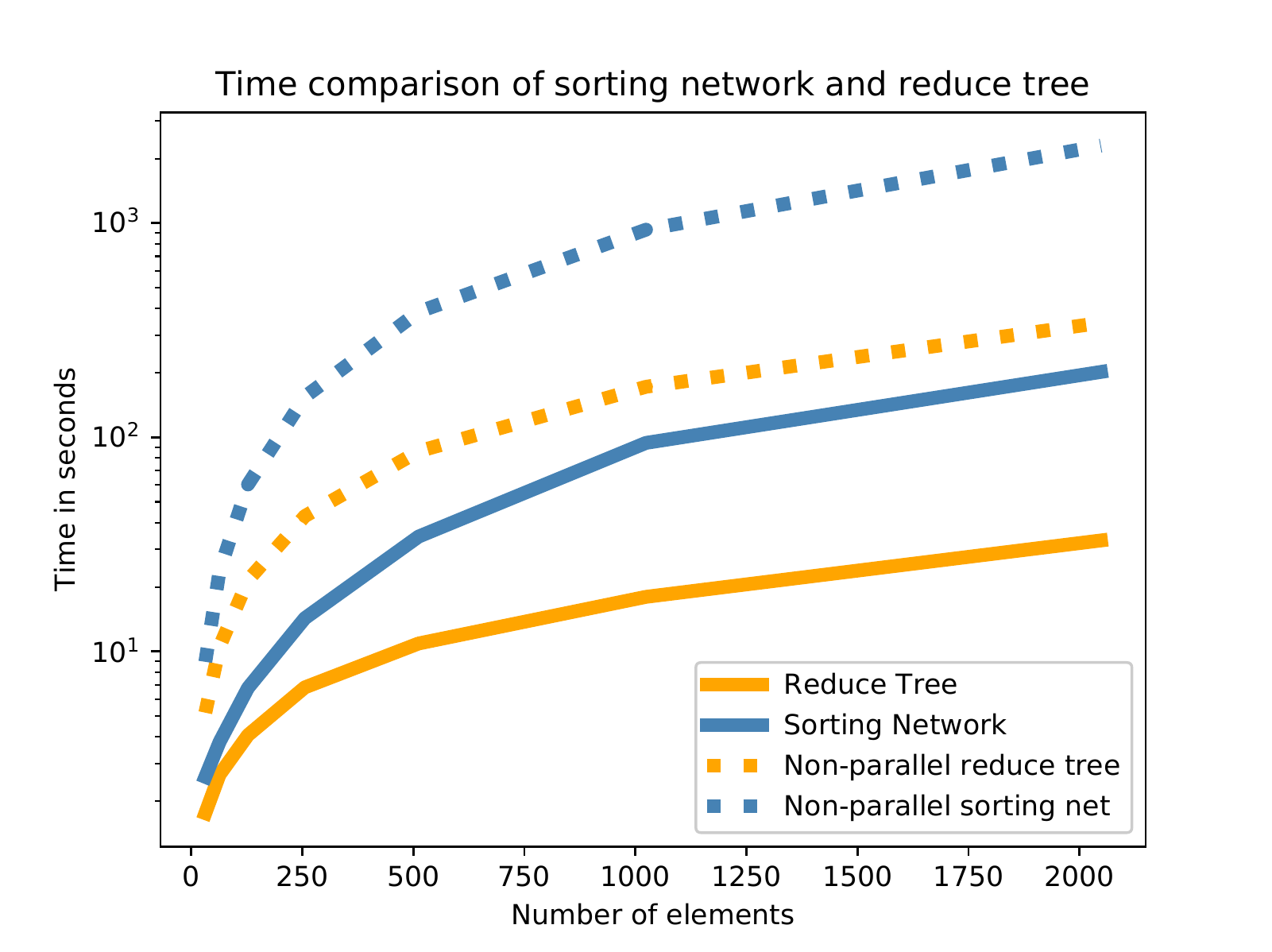}
   \vspace{-5ex}
   \caption{Timing of sorting network and reduce tree addition for different sized vectors, with and without parallelization.}
   \label{Figure.timings}
 \end{figure}

\subsubsection{Batch normalization}
\label{subsubsec:bn}

Batch normalization is mainly used during training; however during evaluating a model this requires us scale and translate and scale the input (which is the output of the previous layer). In practice, when our activation function is the $\sign$ function, this only means that we need to update the bias term (the actual change to the bias term is an elementary calculation). As our circuits are designed to work with a bias term, and the scaling and translation factors are available as plaintext (as they are part of the model), this operation is easily implemented during test time.




\subsection{Sparsification via ``+1''-trick}
\label{ssec:methods-sparsification}

Since we have access to $\mathbf{W} \in \{-1, 1\}^{p \times d}$ and the bias term $\vb \in \integers^p$ in the clear (only data $\x$ and subsequent activations are encrypted), we can exploit the fact that $\mathbf{W}$ always has values $\pm 1$ to roughly halve the cost computation. We consider $\w \in \{-1, 1\}^d$ which is a single row of $\mathbf{W}$ and observe that:
\begin{align}
	\w^\top\x = (\vone + \w)^\top (\vone + \x) - \sum_{i} w_i - (\vone + \x)^\top \vone, \nonumber
\end{align}
where $\vone$ denotes the vector in which every entry is $1$. Further note that
$(\vone + \w) \in \{0,2\}^{d}$ which means that the product $(\vone + \w)^\top
(\vone + \x)$ is simply the quantity $4 \sum_{i: w_i = 1} \bar{x}_i$, where
$\bar{\x}$ refers to the standard binary representation of the non-standard
binary $\x$. Assuming at most half of the $w_i$s were originally $+1$,
if $w \in\{-1, 1\}^d$, only $d/2$ encrypted values need be added. We also need to
compute the encrypted sum $\sum_{i} x_i$; however, this latter sum need only be computed
once, no matter how many output units the layer has. Thus, this small bit of
extra overhead roughly {\em halves} the amount of computation required. We note that
if $\w$ has more $-1$s than $+1$s, $\w^\top \x$ can be computed using $(\vone
- \w)$ and $(\vone - \x)$ instead. This guarantees that we never need to sum
more than half the inputs for any output unit. The sums of encrypted binary
values can be calculated as described in Sec.~\ref{ssec:binaryops}. The
overheads are two additions required to compute $(\vone + \x)^\top \vone$ and
$(\vone - \x)^\top \vone$, and then a subtraction of two $\log(d)$-bit long
encrypted numbers.  (The multiplication by $2$ or $4$ as may be sometimes
required is essentially free, as bit shifts correspond to dropping bits,
and hence do not require homomorphic operations). As our
experimental results show this simple trick roughly halves the computation time
of one layer; the actual savings appear to be even more than half as in many
instances the number of elements we need to sum over is significantly smaller
than half.

It is worth emphasizing the advantage for binarizing and then using the above
approach to making the sums sparse. By default, units in a neural network
compute an affine function to which an activation function is subsequently
applied. The affine map involves an inner product which involves $d$
multiplications. Multiplication under fully homomorphic encryption schemes is
however significantly more expensive than addition. By binarizing and applying
the above calculation, we've replaced the inner product operation by selection
(which is done in the clear as $\mathbf{W}$ is available in plaintext) and
(encrypted) addition.

\subsection{Ternarization (Weight Dropping)}
Ternary neural networks use weights in $\{-1, 0, 1\}$ rather than $\{-1, 1\}$; this can alternatively be viewed as dropping connections from a BNN. Using ternary neural networks rather than binary reduces the computation time as encrypted inputs for which the corresponding $w_i$ is $0$ can be safely dropped from the computation, before the method explained in section~\ref{ssec:methods-sparsification} is applied to the remaining elements. Our experimental results show that a binary network can be ternarized to maintain the same level of test accuracy with roughly a quarter of the weights being $0$ (cf. Sec.~\ref{ssec:accuracy}).

\section{Experimental Results}
\label{sec:results}
In this section we report encrypted binary neural network prediction experiments on a number of real-world datasets. We begin by comparing the efficiency of the two circuits used for inner product, the reduce tree and the sorting network. We then describe the datasets and the architecture of the BNNs used for classification. We report the classification timings of these BNNs for each dataset, for different computational settings. Finally, we give accuracies of the BNNs compared to floating point networks.
Our code is freely available at~\cite{tapas}.

\subsection{Reduce tree vs. sorting network}
We show timings of reduce tree and sorting network for different number of input bits, with and without parallelization in Figure~\ref{Figure.timings} (parallelization is over 16 CPUs). We notice that the reduce tree is strictly better when comparing parallel or non-parallel timings of the circuits. As such, from now on we use the reduce tree circuit for inner product.

It should be mentioned that at the outset this result was not obvious because while sorting networks have more levels of computation, they have fewer gates. Specifically, the sorting network used for encrypted sorting is the bitonic sorting network which for $n$ bits has $O(\log^2 n)$ levels of computation whereas the reduce tree only has $O(\log n)$ levels. On the other hand, the reduce tree requires $2$ gates for each half adder and $5k$ gates for each $k$-bit RC adder, whereas a sorting network only requires $2$ gates per \textsc{SWAP} operation. Another factor that may slow down sorting networks is that is that our implementation of sorting networks is recursive, whereas the reduce tree is iterative.

\subsection{Datasets}
We evaluate on four datasets, three of which have privacy implications due to health care information (datasets Cancer and Diabetes) or applications in surveillance (dataset Faces). We also evaluate on the standard benchmark MNIST dataset.

\paragraph{Cancer.}
The Cancer dataset\footnote{https://tinyurl.com/gl3yhzb} contains $569$ data points where each point has $30$ real-valued features. The task is to predict whether a tumor is malignant (cancerous) or benign. Similar to \citet{MLH:2018} we divide the dataset into a training set and a test in a $70:30$ ratio. For every real-valued feature, we divide the range of each feature into three equal-spaced bins and one-hot encode each feature by its bin-membership. 
This creates a $90$-dimensional binary vector for each example. We use a single fully connected layer $90 \rightarrow 1$ followed by a batch normalization layer, as is common practice for BNNs \cite{CHSEB:2016}.

\paragraph{Diabetes.}
This dataset\footnote{https://tinyurl.com/m6upj7y} contains data on $100000$ patients with diabetes.
The task is to predict one of three possible labels regarding hospital readmission after release. 
We divide patients into a $80/20$ train/test split.
As this dataset contains real and categorical features, we bin them as in the Cancer dataset. We obtain a $1704$ dimensional binary data point for each entry. Our network (selected by cross validation) consists of a fully connected layer $1704 \rightarrow 10$, a batch normalization layer, a \textsc{sign} activation function, followed by another fully connected layer $10 \rightarrow 3$, and a batch normalization layer.

\paragraph{Faces.}
The \textit{Labeled Faces in the Wild-a} dataset contains $13233$ gray-scale face images. We use the binary classification task of gender identification from the images. 
We resize the images to size $50\times 50$. Our network architecture (selected by cross-validation) contains $5$ convolutional layers, each of which is followed by a batch normalization layer and a $\textsc{sign}$ activation function (except the last which has no activation). All convolutional layers have unit stride and filter dimensions $10\times 10$. All layers except the last layer have $32$ output channels (the last has a single output channel). The output is flattened and passed through a fully connected layer $25 \rightarrow 1$ and a batch normalization layer.

\paragraph{MNIST.}
The images in MNIST are $28\times 28$ binary images. The training set and testing sets in this case are already available in a standard split and that is what we use. The training split contains 50000 images and the test split contains 10000 images. There are 10 classes, each corresponding to a different mathematical digit. We use the model (torch7) described in \cite{CHSEB:2016}. 

\subsection{Timing}
\label{sec:parall-strat}
We give timing results for classification of an instance in different computational settings. 
All of the strategies use the parallel implementations of the reduce tree circuit computed across 16 CPUs (the solid line orange line in Figure~\ref{Figure.timings}). The \textit{Out Seq} strategy computes each operation of a BNN sequentially (using the parallel reduce tree circuit). Notice that for any layer of a BNN mapping $d$ inputs to $p$ outputs, the computation over each of the $p$ outputs can be parallelized. 
The \textit{Out 16-P} strategy estimates parallelizing the computation of the $p$ outputs across a cluster of $16$ machines (each with $16$ CPUs). The \textit{Out Full-P} strategy estimates complete parallelization, in which each layer output can be computed independently on a separate machine. We note that for companies that already offer prediction as a service, both of these parallel strategies are not unreasonable requirements. Indeed it is not uncommon for such companies to run hundreds of CPUs/GPUs over multiple days to tune hyperparameters for deep learning models\footnote{https://tinyurl.com/yc8d79oe}. Additionally we report how timings change with the introduction of the \textit{+1-trick} is described in Section \ref{ssec:methods-sparsification}. 
\begin{table}[h!]\centering
  \begin{tabular}[h!]{lrrrr}
  \toprule
    {\bf Parallelism} & {\bf Cancer} & {\bf Diabetes} & {\bf Faces} & {\bf MNIST} \\\hline
    Out Seq & 3.5s & 283s & 763.5h & 65.1h\\\hline
    +1-trick & 3.5s & 250s & 564h & 37.22 h \\\hline
    \begin{tabular}{@{}c@{}} Out 16-P.\\ +1 trick\end{tabular}& 3.5s & 31.5 s & 33.1h&  2.41 h \\\hline
    Out Full-P&3.5s&29s&1.3h& 147s\\
    \bottomrule
  \end{tabular}
  \caption{Neural Network timings on various datasets using different forms of parallelism. \label{tbl:EXP.TIME}}
\end{table}
These timings are given in Table~\ref{tbl:EXP.TIME} (computed with Intel Xeon CPUs @ 2.40GHz, processor number E5-2673V3). We notice that without parallelization over BNN outputs, the predictions on datasets which use fully connected layers: Cancer and Diabetes, finish within seconds or minutes. While the for the datasets that use convolutional layers: Faces and MNIST, predictions require multiple days. The \textit{+1-trick} cuts the time of MNIST prediction by half and reduces the time of Faces prediction by $200$ hours. With only a bit of parallelism over outputs (\textit{Out 16-Parallel}) prediction on the Faces dataset now requires less than 1.5 days and MNIST can be done in $2$ hours. With complete parallelism (\textit{Out N-Parallel}) all methods reduce to under $2$ hours.

\subsection{Accuracy}
\label{ssec:accuracy}
We wanted to ensure that BNNs can still achieve similar test set accuracies to floating point networks. To do so, for each dataset we construct similar floating point networks. For the Cancer dataset we use the same network except we use the original $30$ real-valued features, so the fully connected layer is $30 \rightarrow 1$, as was used in \citet{MLH:2018}. For Diabetes and Faces, just like for our BNNs we cross validate to find the best networks (for Faces: $4$ convolutional layers, with filter sizes of $5 \times 5$ and $64$ output channels; for Diabetes the best network is the same as used in the BNN). For MNIST we report the accuracy of the best performing method \cite{WZZ+:2013} as reported\footnote{https://tinyurl.com/knn2434}. Additionally, we report the accuracy of the weight-dropping method described in Section~\ref{sec:methods}.
\begin{table}[h!]\centering
  \begin{tabular}[h!]{lrrrr}
  \toprule
     & {\bf Cancer} & {\bf Diabetes} & {\bf Faces} & {\bf MNIST} \\
     \hline
    Floating & 0.977 & 0.556 & 0.942 & 0.998  \\\hline
    BNN & 0.971 & 0.549 & 0.891  & 0.986  \\\hline
    \begin{tabular}{@{}l@{}} BNN \\ drop $10\%$ \end{tabular}& 0.976 & 0.549 & 0.879 &  0.976 \\\hline
    \begin{tabular}{@{}l@{}} BNN \\ drop $20\%$ \end{tabular}&0.912 & 0.541 & 0.878 & 0.973\\
    \bottomrule
  \end{tabular}
	\caption{The accuracy of floating point networks compared with BNNs, with and without weight dropping. The Cancer dataset floating point accuracy is given by \cite{MLH:2018}, the MNIST floating point accuracy is given by \cite{WZZ+:2013}, and the MNIST BNN accuracy (without dropping) is given by \cite{CHSEB:2016}.\label{tbl:ACC}}
\end{table}
The results are shown in Table~\ref{tbl:ACC}. We notice that apart from the Faces dataset, the accuracies differ between the floating point networks and BNNs by at most $1.2\%$ (on MNIST). The face dataset uses a different network in floating point which seems to be able to exploit the increased precision to increase accuracy by $5.1\%$. We also observe that weight dropping by $10\%$ reduces the accuracy by at most $1.2\%$ (on Faces). Dropping $20\%$ of the weights seem to have small effect on all datasets except Cancer, which has only a single layer and so likely relies more on every individual weight.



\section{Conclusion}
\label{sec:conclusion}
In this work, we devised a set of techniques that allow for practical Encrypted
Prediction as a Service. In future work, we aim to develop techniques for
encrypting non-binary quantized neural networks, and well as design methods for encrypted model training.






\section*{Acknowledgments}
The authors would like to thank Nick Barlow and Oliver Strickson for their support in using the SHEEP platform. AS acknowledges support from The Alan Turing Institute under the Turing Doctoral Studentship grant TU/C/000023. AG, MK, and VK were supported by The Alan Turing Institute under the EPSRC grant EP/N510129/1.

\bibliography{all-refs}
\bibliographystyle{natbib}

\appendix

\end{document}